\documentclass[conference]{IEEEtran}
\usepackage{cite}
\usepackage{amsmath,amssymb,amsfonts}
\usepackage{algorithmic}
\usepackage{graphicx}
\usepackage{textcomp}
\usepackage{xcolor}
\usepackage{url}
\usepackage[hidelinks]{hyperref}
\def\BibTeX{{\rm B\kern-.05em{\sc i\kern-.025em b}\kern-.08em
    T\kern-.1667em\lower.7ex\hbox{E}\kern-.125emX}}

\begin{document}

\title{Hybrid Quantum-Classical Annealing for Protein Side-Chain Optimization: A Feasibility Study}

\author{
\IEEEauthorblockN{
Valeria B. C. Visani\IEEEauthorrefmark{0},
Simone Reale\IEEEauthorrefmark{0},
Alfonso Gautieri\IEEEauthorrefmark{0},
Elisabetta Di Nitto\IEEEauthorrefmark{0}
}
\IEEEauthorblockA{
\IEEEauthorrefmark{0}\textit{Department of Electronics, Information and Bioengineering, Politecnico di Milano, Milano, Italy}\\
valeriabenedetta.visani@mail.polimi.it, simone1.reale@polimi.it, alfonso.gautieri@polimi.it, elisabetta.dinitto@polimi.it
}
}

\maketitle

\begin{abstract}
Hybrid quantum-classical annealing solvers are moving from research laboratories toward industrial use, but their advantage over classical optimization appears to be confined to a narrow class of problems. A recent systematic benchmark of the D-Wave hybrid solver against industry-leading classical solvers, across several classes of optimization problems, found it to be most competitive on binary quadratic programs, while noting that few industrial problems take this form. Therefore, identifying real-world industrial problems that can be modeled with this structure is a key challenge. We propose protein side-chain optimization, a core step of computational protein design, as one of these problems: it is natively a binary quadratic program, and its application context requires pools of low-energy solutions, which annealers naturally return. In this feasibility study, we test whether the competitiveness of the hybrid solver against industry-leading classical solvers extends to this problem. On 40 instances derived from 20 experimentally determined protein structures with up to about 200 amino acids and 2000 binary variables, we compare the results achieved with a hybrid solver, run for 5 seconds, against those of the commercial solver Gurobi, run for 10 seconds, 1000 seconds, and 4 hours. The hybrid solver reaches the certified optimum whenever one is available, matches Gurobi's performance on smaller instances, and outperforms it on larger instances, finding lower-energy solutions. These results show that the competitiveness of the hybrid solver on binary quadratic programs, so far observed only on a synthetic benchmark, extends to a real-world problem. This motivates comparisons with classical heuristics specific to protein side-chain optimization.
\end{abstract}

\section{Introduction}
\label{sec:intro}

Quantum annealing is an analog computing paradigm in which an optimization problem is encoded into the physical Hamiltonian of the processor. Current annealers implement only a restricted class of Hamiltonians, so they are not universal quantum computers; however, they have the advantage of being more mature than general-purpose gate-based quantum devices: D-Wave processors have grown from 512 qubits in 2013 to more than 4,400 qubits in the current Advantage2 systems \cite{dwave_advantage2_datasheet_2025}. In addition, hybrid quantum-classical solvers combine both types of computation: they are built on classical algorithms and call the Quantum Processing Unit (QPU) only when it is useful. Since the quantum component is added on top of a classical solver, these systems are already approaching practical maturity and are moving from research to industrial applications \cite{Quinton2025}.

Their practical computational advantage, however, appears to be limited to a narrow class of problems. A systematic comparison with commercial classical solvers on linear, quadratic, and mixed-integer formulations \cite{Quinton2025} found the D-Wave hybrid solvers to be most competitive on Binary Quadratic Programs (BQPs), i.e., problems that are natively quadratic in binary variables, while noting that only a limited number of industrial optimization problems are pure BQPs. Identifying practically relevant problems with this structure is therefore a prerequisite for the effective application of hybrid annealing.

In this paper, we propose a real-world industrial problem that falls into this class: Protein Side-Chain Optimization (PSCO), described in Sec.~\ref{sec:psco}. PSCO is a core step of computational protein design, and is solved routinely in industrial and academic design pipelines within Rosetta, one of the most widely used software suites for molecular modeling, through a simulated annealing algorithm \cite{rosetta2020}. With the energy functions used in practice, its objective is exactly quadratic in binary selection variables, and its only constraints are one-hot constraints on groups of these variables: PSCO is then natively a BQP. In addition, protein design requires pools of many low-energy candidates rather than a single optimum, which matches the ability of annealers to return samples of low-energy solutions.

Quantum approaches to PSCO have been explored with several quantum technologies. The first study based on quantum annealing \cite{qPacker2019} analyzed how the time to solution scales with problem size on design problems small enough to be solved directly on the D-Wave 2000Q processor, with up to about 60 binary variables, and compared it with the Rosetta packer and the exact solver toulbar2; under pessimistic assumptions on future connectivity and noise, the authors estimated that annealers about 400 and 10,000 times larger would be needed to compete with toulbar2 and with the Rosetta packer, respectively. The scaling comparison therefore concerned the quantum processor alone. The hybrid algorithm qbsolv was used only on larger instances to assess solution quality without a scaling analysis. Recent work has applied photonic entropy-computing hardware to PSCO instances involving 493 to 943 variables \cite{Emami2026}, while a separate study explored the Quantum Approximate Optimization Algorithm (QAOA) on gate-based quantum computers for the same problem up to 54 qubits \cite{Agathangelou2025}. Quantum annealing, either directly on the QPU or through D-Wave hybrid solvers, has also been applied to other problems in protein modeling, including folding, design, and conformational sampling \cite{Irback2022, irback2024, Meuser2025, Panizza2024, Tucs2023, Ghamari2024}. Two of these studies compare a D-Wave hybrid solver with Gurobi, finding Gurobi to perform better \cite{Panizza2024} or comparably \cite{Meuser2025}. A systematic comparison between current hybrid solvers and classical solvers on PSCO instances of increasing, realistic size, which would show whether the competitiveness found in \cite{Quinton2025} also holds for this problem, is still missing.

In this paper, we address this question as a feasibility study, to assess whether this direction is worth a more detailed investigation. Our contributions are the following:
\begin{itemize}
\item We propose PSCO as a real-world industrial problem suited to hybrid annealing, since it is natively a BQP with only one-hot constraints and its application context requires pools of low-energy solutions, which annealers naturally return.
\item We formulate PSCO, with energies computed by Rosetta on experimentally determined protein structures, as a constrained quadratic model that the hybrid solver receives with explicit one-hot constraints.
\item We benchmark the D-Wave hybrid solver, run for 5~s, against Gurobi, run for 10~s and 1000~s, on 40 instances with up to 1970 binary variables, using as reference 4~h Gurobi runs that certify optimality on the smallest instances.
\item We show how the quality gap between the two solvers evolves with instance size, and discuss what this result does and does not imply with respect to classical heuristics specific to PSCO.
\end{itemize}

\section{Protein Side-Chain Optimization}
\label{sec:psco}

Proteins are polypeptide chains of amino acids, each of which is referred to as a \emph{residue} within the chain. Typical proteins contain hundreds of residues. The 20 canonical amino acids share a common set of atoms, which together form the \emph{backbone} of the chain, but differ in their side chains, which contribute to the structure and physicochemical properties of the protein. In computational modeling, side-chain conformations are commonly represented by discrete states called \emph{rotamers}, which sample preferred combinations of side-chain torsion angles.

Protein engineering aims to modify proteins to obtain desired properties, often by replacing amino acids at selected positions \cite{Zhou2024}. Allowing all 20 canonical amino acids at $k$ positions generates $20^k$ possible sequences, making exhaustive experimental evaluation impractical even for relatively small $k$. Computational design therefore helps prioritize candidates for experimental testing, often through successive stages of structural optimization and filtering \cite{Zhou2024}.

PSCO is a central component of this process. Given a backbone with $N$ residues and a set of $n$ rotamers per residue, PSCO selects the combination of rotamers that minimizes the total energy of the protein among the $n^N$ possible conformations \cite{Gainza2016}. This problem has been shown to be NP-hard \cite{Pierce2002}.

When the allowed rotamers at each position belong to a single, prescribed amino acid type, PSCO optimizes side-chain conformations for a fixed sequence, a task commonly termed \textit{side-chain repacking}. When multiple amino acid types are allowed, selecting a rotamer determines both the amino acid identity and its side-chain conformation; this joint optimization of sequence and side-chain conformations is referred to as \textit{fixed-backbone protein design}~\cite{Gainza2016}. The present study addresses this design setting using restricted sets of allowed rotamers, with the minimum identifying energetically favorable candidate designs.

\section{Why PSCO for hybrid annealing}

Two properties make PSCO a natural candidate for hybrid annealing. First, as noted above, PSCO is natively a BQP whose only constraints are one-hot constraints, i.e., linear equality constraints requiring exactly one variable in a group of binary variables to be equal to one (here, exactly one rotamer per residue). Consequently, it belongs to the class on which hybrid solvers were found to be most competitive \cite{Quinton2025}. Second, the early stages of protein engineering require pools of many diverse low-energy candidates rather than a single optimum, since these candidates are then evaluated in later stages with more accurate filters \cite{Zhou2024}. Annealers are well suited to this task, as they return a large and diverse set of good-quality solutions rather than only the global optimum \cite{Pang2021, Mohseni2023}. This work tests the first property; whether hybrid annealing also provides an advantage in sampling diverse near-optimal solutions is left for future work.

\section{Formulation}
\label{sec:formulation}

Quantum annealers minimize objective functions in Quadratic Unconstrained Binary Optimization (QUBO) form. Given a vector of $M$ binary variables $\mathbf{x} \in \{0,1\}^M$ and a symmetric matrix $Q \in \mathbb{R}^{M \times M}$, the objective is
\begin{equation}
f_Q(\mathbf{x}) = \sum_{a=1}^{M} Q_{aa}\, x_a + \sum_{a<b} Q_{ab}\, x_a x_b,
\label{eq:qubo}
\end{equation}
where the diagonal entries $Q_{aa}$ act as linear terms because $x_a^2 = x_a$. In this form, constraints can only be imposed through penalty terms added to the objective. 

To solve PSCO using hybrid annealing, we express its energy as an objective of the form~\eqref{eq:qubo} and its constraints as linear equalities on the same variables.
More specifically,
we adopt a one-hot encoding, associating a binary variable $x_{i,r}$ with each possible rotamer choice for each residue. The variable equals one if rotamer $r$ is selected for residue $i$ and zero otherwise. Thus, an instance with $N$ residues and $n$ rotamers has $M = N\times n$ variables. With a pairwise-decomposable energy function, the total energy is
\begin{equation}
E(\mathbf{x}) = \sum_{i=1}^{N}\sum_{r=1}^{n} E_{i,r}\, x_{i,r}
 + \sum_{i<j}\sum_{r,s} E_{i,r;j,s}\, x_{i,r}\, x_{j,s},
\label{eq:energy}
\end{equation}
where the one-body terms $E_{i,r}$ represent the energy contributions of individual residues adopting specific rotamers, and the two-body terms $E_{i,r;j,s}$ represent the interaction energies between pairs of residues adopting the corresponding rotamers. These terms are nonzero only for residues that can interact. Equation~\eqref{eq:energy} has exactly the form of~\eqref{eq:qubo}, with the one-body terms on the diagonal of $Q$ and the two-body terms off the diagonal. Since only nearby residues interact, $Q$ is sparse for large proteins.

The only constraints of PSCO are the one-hot constraints
\begin{equation}
\sum_{r=1}^{n} x_{i,r} = 1, \qquad i = 1,\dots,N,
\label{eq:onehot}
\end{equation}
which ensure that exactly one rotamer is selected for each residue. Since the number of rotamers $n$ is an input parameter, a residue may have fewer available rotamers $n_i < n$. In that case, the variables of the missing rotamers are fixed by the additional constraints
\begin{equation}
x_{i,r} = 0, \qquad r > n_i,
\label{eq:zero}
\end{equation}
so that the model remains a BQP with linear equality constraints. Since these variables are fixed to zero, the problem is equivalent to one without them, whose only constraints are the one-hot constraints~\eqref{eq:onehot} over the available rotamers.

The hybrid solver receives~\eqref{eq:energy}--\eqref{eq:zero} as a constrained quadratic model (CQM) \cite{dwaveCQM2021}, with all constraints stated explicitly. No penalty weights need to be specified by the user, as any penalty encoding is handled internally by the solver. Gurobi receives the same objective and constraints as a binary quadratic programming (BQP) model with linear equality constraints, so that both solvers address the same mathematical problem.

\section{Methods}
\label{sec:methods}

\subsection*{Problem Instances}
We retrieved 20 protein structures from the RCSB Protein Data Bank (PDB) \cite{pdb2000}; the complete list is available in our repository (see Section~\nameref{sec:data}). The side chains were removed from each structure with the molecular modeling software PyMOL \cite{pymol}. The one-body and two-body energy terms were computed with PyRosetta \cite{pyrosetta2010} with the \texttt{ref2015} energy function \cite{ref2015}, in Rosetta Energy Units (REU), using the default packer task, which allows every amino acid type at every position. To keep the problem tractable, for each residue we selected a subset of 5 or 10 consecutive rotamers of the set generated by Rosetta in design mode. This yields 40 problem instances ranging from 50 to 1970 binary variables.

\subsection*{Compared Methods and Reference Values}
\paragraph*{Hybrid annealing (HA)} We used the D-Wave Leap hybrid solver for constrained quadratic models (version 1.14)\cite{dwaveCQM2021}, which runs classical heuristics on cloud servers and queries a QPU on subproblems \cite{dwaveHSS2022}; neither the heuristics nor the classical hardware are disclosed. The solver receives the objective defined in Sec.~\ref{sec:formulation} with the constraints stated explicitly. The only user parameter is the time limit, which we set to 5~s, the minimum possible value.

\paragraph*{Gurobi} We used Gurobi (version 13.0.1) \cite{gurobi2025} in two configurations, both with 16 threads on an Intel Xeon Silver 4114 (16 vCPUs, 2.2~GHz, 62~GiB of RAM). In the \emph{default configuration}, used to compute the reference values, Gurobi runs until it certifies the optimum or reaches a time limit, set to 4~h, beyond which we observed no significant improvement. In the \emph{custom configuration}, used for comparison and intended to find good solutions quickly, we set \texttt{MIPFocus}~=~1, \texttt{Heuristics}~=~0.5, and \texttt{MIPGap}~=~0.1, the latter stopping the solver once the relative gap to its bound falls below 10\%; we used time limits of 10~s, the same order of magnitude as the HA time, and 1000~s, as in \cite{Quinton2025}.

 \subsection*{Experimental Setup and Evaluation Metric}
For each of the 40 instances, we performed one Gurobi run in the default configuration, five Gurobi runs in the custom configuration for each time limit, using different random seeds, and five independent HA runs. The reference energy $E_{\mathrm{ref}}$ is the energy found by the default Gurobi run. It is a certified optimum for instances on which Gurobi proved optimality, i.e. up to 60 residues with 5 rotamers and up to 30 residues with 10 rotamers. For the remaining instances, it is the best solution found within 4~h, which can be improved upon by other runs. For each instance and method, we compute the maximum, minimum, and mean energy, and calculate the percentage gap $(E - E_{\mathrm{ref}})/|E_{\mathrm{ref}}|$ for each of these energy values relative to the reference. Negative gaps indicate solutions with lower energy than the reference.

\section{Results}
\label{sec:results}

\begin{figure*}[t]
  \centering
  \includegraphics[width=\textwidth]{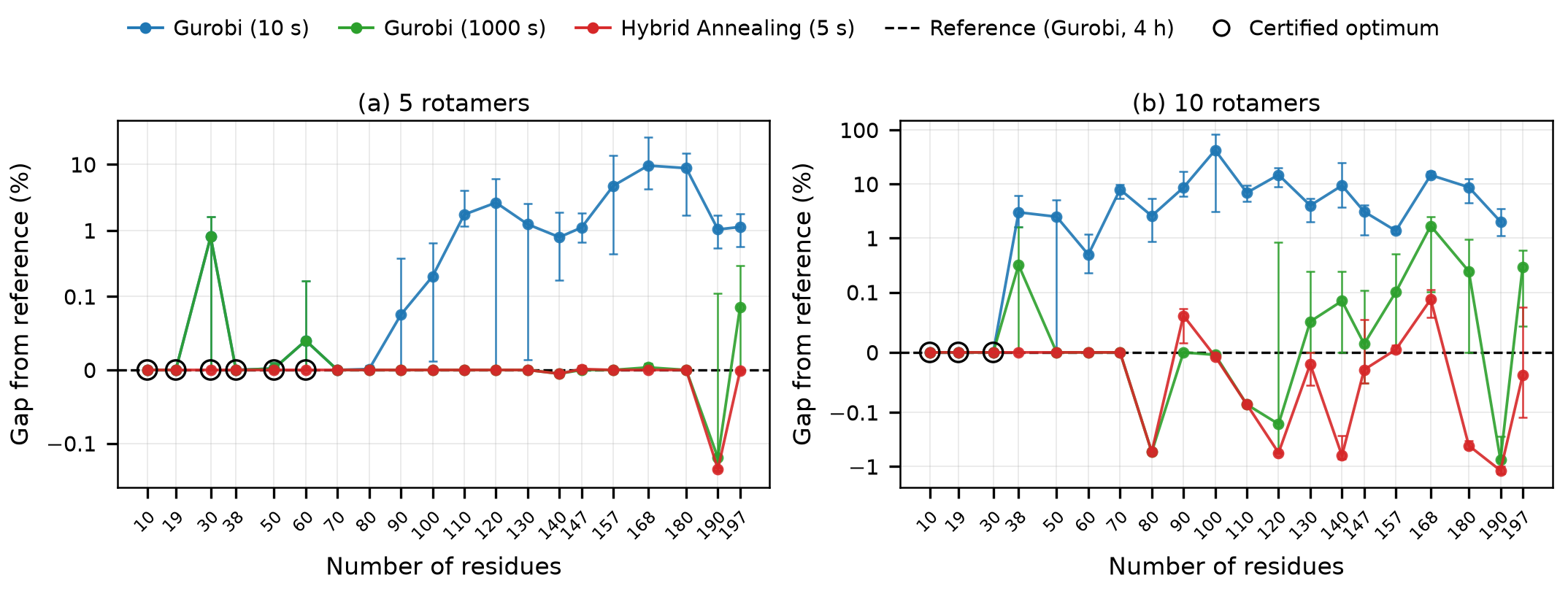}
  \caption{Percentage gap $(E - E_{\mathrm{ref}})/|E_{\mathrm{ref}}|$ from the reference for instances from 10 residues to 197 residues with (a)~5 rotamers and (b)~10 rotamers. The reference (dashed line) is the best energy found by Gurobi in its default configuration within 4~h. It is compared with Hybrid Annealing with a 5~s budget and with Gurobi in its custom configuration with budgets of 10~s and 1000~s. Markers show the mean over 5 runs, and error bars the minimum and maximum. In both panels the vertical axis is linear within $\pm 0.1\%$ and logarithmic beyond; negative values indicate energies below the reference. Note the different vertical-axis ranges in the two panels. Circles mark the instances where the reference is a certified optimum. Gurobi with a 10~s budget is omitted for the instance with 197 residues and 10 rotamers, as it did not return a meaningful solution within the time limit.}
  \label{fig:quality}
\end{figure*}

\paragraph*{5 rotamers per residue} Figure \ref{fig:quality}(a) reports the results for instances with 10 to 197 residues considering 5 rotamers per residue. HA reaches the reference on every instance of up to 180 residues, including the certified optima up to 60 residues, and finds slightly lower energies than the 4~h Gurobi default configuration solutions on the two largest instances, by 2.8~REU (0.25\%) at 190 residues and by 0.04~REU at 197 residues. Apart from three instances discussed below, Gurobi custom configuration with a 10 s budget deviates from the reference from 90 residues onward, with mean gaps of 0.8--9.6\% from 110 residues onward. With a 1000 s budget, Gurobi custom configuration matches HA up to 180 residues, with the same three exceptions, but is worse on the two largest instances, by about 0.9~REU at 190 residues and 3.5~REU at 197 residues. The exceptions are the instances with 30, 50, and 60 residues, on which Gurobi runs stop within a few seconds at solutions up to 2.1~REU above the certified optimum, with identical results for both budgets, because the 10\% gap criterion is met before the optimum is reached.

\paragraph*{10 rotamers per residue} Figure \ref{fig:quality}(b) reports the results for instances with 10 to 197 residues considering 10 rotamers per residue. Gurobi default configuration 4 h run certifies optimality only up to 30 residues, and HA reaches the optimum on these instances. From 80 residues onward, HA finds lower energies than the 4 h Gurobi solution on 10 of the 13 instances, by up to 8.0~REU (1.2\%) at 190 residues; on the remaining three (90, 157, and 168 residues), its mean energy difference is between 0.04 and 0.38~REU. Gurobi custom configuration with a 10 s budget is far from the reference from 38 residues onward, with mean gaps of 0.5--15\% and a peak of 42\% at 100 residues. With a 1000 s budget, Gurobi custom configuration matches HA up to 110 residues, with HA 0.08~REU above Gurobi at 90 residues, and falls behind it on all instances from 120 residues onward, by 0.2 to 6.8~REU in mean energy. Gurobi with a 10~s budget is omitted for the instance with 197 residues, as it did not return a meaningful solution within the time limit.

\paragraph*{Scaling and resources} On both groups of instances, the advantage of HA over Gurobi grows with problem size, and it appears earlier with more rotamers: with 5 rotamers, HA matches Gurobi with a 1000~s budget up to about 180 residues and improves on it only beyond, whereas with 10 rotamers the crossover occurs at about 110 residues, after which HA also finds energies below the 4~h reference on most instances. In both cases the crossover corresponds to roughly $10^3$ binary variables. Two considerations apply to this comparison. First, the classical components of the hybrid solver, including both the algorithms and the hardware on which they run, are not publicly disclosed, as stated above. Second, the QPU access time is only about 0.63\% of the HA runtime and does not vary with the problem size. This implies that the quality of our HA results cannot be attributed to the quantum component alone.

\section{Discussion and Conclusions}
\label{sec:conclusions}

On protein side-chain optimization instances derived from real structures, the best of five 5~s runs of the D-Wave hybrid solver matches or improves on the best solution found by Gurobi (10~s, 1000~s and 4~h) on 37 of 40 instances, including all certified optima, and on five of the larger instances it finds the lowest energy obtained by any method; on the remaining three, its best energy is at most 0.06\% higher than Gurobi's best. These results support the hypothesis of Sec.~\ref{sec:intro}: the competitiveness of hybrid annealing found in \cite{Quinton2025} on BQPs extends to a practically relevant problem that is natively in this form, without reformulation and without user-defined penalty weights.

This work therefore identifies a candidate near-term industrial application of hybrid annealing: a real-world problem whose mathematical structure matches the class on which hybrid solvers are competitive, and on which, at realistic sizes, the hybrid solver already outperforms an industry-leading general-purpose solver. As a preliminary feasibility study, however, it does not establish an advantage over the best classical methods for PSCO, and, since the classical components of the hybrid solver are undisclosed and only a small fraction of its runtime is spent on the QPU, its results cannot be attributed to the quantum processor alone.

These results motivate two subsequent studies. First, the hybrid solver can be compared with a simulated annealing algorithm optimized for PSCO, such as the one used in Rosetta, for two reasons: \textit{i.} to verify whether its advantage persists against a problem-specific classical heuristic on instances of increasing size; \textit{ii.} to compare the ability of the two approaches to provide pools of diverse near-optimal candidate solutions. Second, a problem-specific custom hybrid solver could be developed to isolate the contribution of the QPU in the solution of our protein engineering problem.

\section*{Data and Code Availability}
\phantomsection
\label{sec:data}

Data and code associated with this manuscript are publicly available at:
\textit{\url{https://github.com/valeriavisani/hybrid_annealing_in_protein_engineering_feasibility}}.

\section*{Acknowledgments}
We acknowledge the CINECA award under the ISCRA initiative, for the availability of quantum computing resources and support.


\begin{thebibliography}{00}

\bibitem{dwave_advantage2_datasheet_2025}
D-Wave Quantum Inc., \textit{D-Wave Advantage2 System: Production-Ready Quantum Performance for Business and Science}, Data Sheet, May 2025. [Online]. Available: \url{https://www.dwavequantum.com/media/3ikjqkyw/d-wave-advantage2-data-sheet-final.pdf}

\bibitem{Quinton2025}
F. A. Quinton, P. A. S. Myhr, M. Barani, P. Crespo del Granado, and H. Zhang, ``Quantum annealing applications, challenges and limitations for optimisation problems compared to classical solvers", \textit{Scientific Reports}, vol. 15, Art. no. 12733, 2025, doi: \href{https://doi.org/10.1038/s41598-025-96220-2}{10.1038/s41598-025-96220-2}.

\bibitem{rosetta2020}
J. K. Leman, B. D. Weitzner, S. M. Lewis, et al., ``Macromolecular modeling and design in Rosetta: recent methods and frameworks", \textit{Nature Methods}, vol. 17, no. 7, pp. 665--680, 2020, doi: \href{https://doi.org/10.1038/s41592-020-0848-2}{10.1038/s41592-020-0848-2}.

\bibitem{qPacker2019}
V. K. Mulligan, H. Melo, H. I. Merritt, S. Slocum, B. D. Weitzner, A. M. Watkins, P. D. Renfrew, C. Pelissier, P. S. Arora, and R. Bonneau, ``Designing Peptides on a Quantum Computer", bioRxiv preprint, 2019, doi: \href{https://doi.org/10.1101/752485}{10.1101/752485}.

\bibitem{Emami2026}
B. Emami, W. Dyk, D. Haycraft, J. Robinson, L. Nguyen, M.-A. Miri, and D. J. Huggins, ``Entropy quantum computing for fixed-backbone protein design", \textit{Scientific Reports}, vol. 16, Art. no. 26710, 2026, doi: \href{https://doi.org/10.1038/s41598-026-54101-2}{10.1038/s41598-026-54101-2}.

\bibitem{Agathangelou2025}
A. Agathangelou, D. Manawadu, and I. Tavernelli, ``Quantum Algorithm for Protein Side-Chain Optimisation: Comparing Quantum to Classical Methods", arXiv preprint arXiv:2507.19383, 2025, doi: \href{https://doi.org/10.48550/arXiv.2507.19383}{10.48550/arXiv.2507.19383}.

\bibitem{Panizza2024}
V. Panizza, P. Hauke, C. Micheletti, and P. Faccioli, ``Protein Design by Integrating Machine Learning and Quantum-Encoded Optimization", \textit{PRX Life}, vol. 2, no. 4, Art. no. 043012, 2024, doi: \href{https://doi.org/10.1103/PRXLife.2.043012}{10.1103/PRXLife.2.043012}.

\bibitem{Meuser2025}
L. Meuser, A. Patsilinakos, and P. Faccioli, ``De Novo Design of Protein-Binding Peptides by Quantum Computing", \textit{Journal of Chemical Theory and Computation}, vol. 21, no. 19, pp. 9993--10005, 2025, doi: \href{https://doi.org/10.1021/acs.jctc.5c00768}{10.1021/acs.jctc.5c00768}.

\bibitem{irback2024}
A. Irbäck, L. Knuthson, S. Mohanty, and C. Peterson, ``Using quantum annealing to design lattice proteins", \textit{Physical Review Research}, vol. 6, no. 1, Art. no. 013162, 2024, doi: \href{https://doi.org/10.1103/PhysRevResearch.6.013162}{10.1103/PhysRevResearch.6.013162}.

\bibitem{Irback2022}
A. Irb\"{a}ck, L. Knuthson, S. Mohanty, and C. Peterson, ``Folding lattice proteins with quantum annealing", \textit{Physical Review Research}, vol. 4, no. 4, Art. no. 043013, 2022, doi: \href{https://doi.org/10.1103/PhysRevResearch.4.043013}{10.1103/PhysRevResearch.4.043013}.

\bibitem{Tucs2023}
A. Tu\v{c}s, F. Berenger, A. Yumoto, R. Tamura, T. Uzawa, and K. Tsuda, ``Quantum annealing designs nonhemolytic antimicrobial peptides in a discrete latent space", \textit{ACS Medicinal Chemistry Letters}, vol. 14, no. 5, pp. 577--582, 2023, doi: \href{https://doi.org/10.1021/acsmedchemlett.2c00487}{10.1021/acsmedchemlett.2c00487}.

\bibitem{Ghamari2024}
D. Ghamari, R. Covino, and P. Faccioli, ``Sampling a rare protein transition using quantum annealing", \textit{Journal of Chemical Theory and Computation}, vol. 20, no. 8, pp. 3322--3334, 2024, doi: \href{https://doi.org/10.1021/acs.jctc.3c01174}{10.1021/acs.jctc.3c01174}.

\bibitem{Zhou2024}
J. Zhou and M. Huang, ``Navigating the landscape of enzyme design: from molecular simulations to machine learning", \textit{Chemical Society Reviews}, vol. 53, no. 16, pp. 8202--8239, 2024, doi: \href{https://doi.org/10.1039/D4CS00196F}{10.1039/D4CS00196F}.

\bibitem{Gainza2016}
P. Gainza, H. M. Nisonoff, and B. R. Donald, ``Algorithms for protein design", \textit{Current Opinion in Structural Biology}, vol. 39, pp. 16--26, 2016, doi: \href{https://doi.org/10.1016/j.sbi.2016.03.006}{10.1016/j.sbi.2016.03.006}.

\bibitem{Pierce2002} N. A. Pierce and E. Winfree, ``Protein design is NP-hard'',
\emph{Protein Eng.}, vol. 15, no. 10, pp. 779--782, 2002, doi: \href{https://doi.org/10.1093/protein/15.10.779}{10.1093/protein/15.10.779}

\bibitem{Pang2021}
Y. Pang, C. Coffrin, A. Y. Lokhov, and M. Vuffray, ``The potential of quantum annealing for rapid solution structure identification'', \textit{Constraints}, vol. 26, no. 1, pp. 1--25, 2021, doi: \href{https://doi.org/10.1007/s10601-020-09315-0}{10.1007/s10601-020-09315-0}.

\bibitem{Mohseni2023}
M. Mohseni, M. M. Rams, S. V. Isakov, D. Eppens, S. Pielawa, J. Strumpfer, S. Boixo, and H. Neven, ``Sampling diverse near-optimal solutions via algorithmic quantum annealing", \textit{Physical Review E}, vol. 108, Art. no. 065303, 2023, doi: \href{https://doi.org/10.1103/PhysRevE.108.065303}{10.1103/PhysRevE.108.065303}.

\bibitem{dwaveCQM2021}
D-Wave Systems Inc., \textit{Hybrid Solver for Constrained Quadratic Models}, White Paper, no. 14-1055A-A, 2021. [Online]. Available: \url{https://www.dwavesys.com/media/rldh2ghw/14-1055a-a_hybrid_solver_for_constrained_quadratic_models.pdf}

\bibitem{pdb2000}
H. M. Berman, J. Westbrook, Z. Feng, et al., ``The Protein Data Bank", \textit{Nucleic Acids Research}, vol. 28, no. 1, pp. 235--242, 2000, doi: \href{https://doi.org/10.1093/nar/28.1.235}{10.1093/nar/28.1.235}.

\bibitem{pymol}
Schr\"odinger, LLC, \textit{The PyMOL Molecular Graphics System}. [Online]. Available: \url{https://www.pymol.org}

\bibitem{pyrosetta2010}
S. Chaudhury, S. Lyskov, and J. J. Gray, ``PyRosetta: a script-based interface for implementing molecular modeling algorithms using Rosetta", \textit{Bioinformatics}, vol. 26, no. 5, pp. 689--691, 2010, doi: \href{https://doi.org/10.1093/bioinformatics/btq007}{10.1093/bioinformatics/btq007}.

\bibitem{ref2015}
R. F. Alford, A. Leaver-Fay, J. R. Jeliazkov, M. J. O'Meara, F. P. DiMaio, H. Park, M. V. Shapovalov, P. D. Renfrew, V. K. Mulligan, K. Kappel, J. W. Labonte, M. S. Pacella, R. Bonneau, P. Bradley, R. L. Dunbrack, R. Das, D. Baker, B. Kuhlman, T. Kortemme, and J. J. Gray, ``The Rosetta All-Atom Energy Function for Macromolecular Modeling and Design", \textit{Journal of Chemical Theory and Computation}, vol. 13, no. 6, pp. 3031--3048, 2017, doi: \href{https://doi.org/10.1021/acs.jctc.7b00125}{10.1021/acs.jctc.7b00125}.

\bibitem{dwaveHSS2022}
D-Wave Systems Inc., \textit{D-Wave Hybrid Solver Service: An Overview}, White Paper, 2022. [Online]. Available: \url{https://www.dwavequantum.com/resources/white-paper/d-wave-hybrid-solver-service-an-overview/}

\bibitem{gurobi2025}
Gurobi Optimization, LLC, \textit{Gurobi Optimizer Reference Manual}, 2025. [Online]. Available: \url{https://www.gurobi.com}

\end{thebibliography}
\end{document}